\documentclass[prd,eqsecnum,aps, showpacs]{revtex4}
\usepackage{epsf,graphics,graphicx}
\usepackage{amsmath}
\usepackage{amssymb,latexsym,mathrsfs}
\usepackage{hyperref}

\def\bea{\begin{eqnarray}}
\def\eea{\end{eqnarray}}
\def\ba{\begin{array}}
\def\ea{\end{array}}

\def\beq{\begin{equation}}
\def\eeq{\end{equation}}

\begin{document}
%\title{Quantum fluctuations in mutated hilltop inflation}
%\title{Quantum fluctuations and observational aspects of a scalar field driven inflation}
\title{A semi-analytical approach to perturbations in mutated hilltop inflation}

\author{Barun Kumar Pal$^{1}$\footnote{Electronic address: {barunp1985@rediffmail.com}}, ${}^{}$ Supratik Pal$^{1, 2}$\footnote{Electronic address: {pal@th.physik.uni-bonn.de}} ${}^{}$
and B. Basu$^{1}$\footnote{Electronic address: {banasri@isical.ac.in}} ${}^{}$}
\affiliation{$^1$Physics and Applied Mathematics Unit, Indian Statistical Institute, 203 B.T. Road, Kolkata 700 108, India \\
$^2$Bethe Center for Theoretical Physics and Physikalisches Institut der Universit\"{a}t Bonn, Nussallee 12, 53115 Bonn, Germany}

\vspace{1in}

\begin{abstract}
We study cosmological perturbations and observational aspects for mutated hilltop model of inflation. Employing mostly analytical treatment,
we evaluate observable parameters during inflation as well as post-inflationary perturbations.
This further leads to exploring observational aspects related to Cosmic Microwave Background (CMB) radiation. This semi-analytical treatment reduces complications related to numerical computation to some extent for studying the different phenomena related to CMB angular power spectrum for mutated hilltop inflation.
\end{abstract}


%\vspace{1in}
\maketitle
%%%%%%%%%%%%%%%%%%%%%%%%%%%%%%%%%%%%%%%%%%%%%%%%%%%%%%%%%%%%%%%%%%%%%%%%%%%%%%%%%%%%%%%%%%%%%%%%%%%%%%%%%%%%%%%%%%%%%%%%%%%%%%%%%%%%%%%%%%%%%%%%%%%%%%%%%
%%%%%%%%%%%%%%%%%%%%%%%%%%%%%%%%%%%%%%%%%%%%%%%%%%%%%%%%%%%%%%%%%%%%%%%%%%%%%%%%%%%%%%%%%%%%%%%%%%%%%%%%%%%%%%%%%%%%%%%%%%%%%%%%%%%%%%%%%%%%%%%%%%%%%%%%%
\section{introduction}
The most striking aspect of cosmological inflationary scenario \cite{guth} is that they provide natural explanation
for the origin of cosmological perturbation seeds for structure formation leading to a nearly scale-free
power spectrum which is in tune with direct  observational evidence from Cosmic Microwave Background Radiation (CMB) \cite{wmap3,wmap07}
and Large Scale Structure Formation \cite{large}. The form of the primordial perturbations seems to explain the
observed temperature anisotropy in the CMB \cite{wmap3,wmap07}.

%% 5555 As the last two decades have seen giant strides in observational cosmology, any successful model of inflation has to be compatible with
%the precise observational data from cosmic microwave background (CMB)\cite{penzias,julian67,wmap07} or other independent measures \cite{jles07}.
%All the existing models have some positive features or the other, but the picture is still not completely satisfactory. As of now the debate on the candidate for {\it inflaton} is still ongoing.
Recently we have proposed a variant of  hilltop inflation, namely, mutated hilltop inflation \cite{barun}, in order to explain the
early universe. The analytical expressions for most of the observable parameters have been obtained. However, as pointed out in that paper, the expression for the ratio of the tensor to scalar amplitudes does not \textit{exactly} conform with the usual consistency relation $r = -8 n_T$ \cite{andrew}. In the present article, we have tried to search for a possible explanation for the same by deviating slightly from perfect de-Sitter approximation right from the beginning.
This turns out to be a complimentary approach of the previously employed approaches of obtaining modified consistency relation, such as, when higher order terms  of the slow roll parameters are taken into account \cite{stewart,0101225,PRD66}, if {\it tensor to scalar ratio} is expressed as a function of {\it tensor spectral index, scalar spectral index} and {\it running of the tensor spectral index} \cite{parker}, if generalized propagation speed (less than one) of the scalar field fluctuations relative to the homogeneous background is considered \cite{peiris,bean,mukhanov} and the likes.

The advent of highly precise data from the observational probes like COBE \cite{cobe}, WMAP \cite{wmap3} and the
forthcoming Planck \cite{planck} has made the task for the development of the framework of post-inflationary perturbation
more difficult in order to met up with challenges both from building up a theoretical framework of post-inflationary
perturbative technique and from confronting the theory with observations more precisely. This usually leads to
employ  direct numerical techniques by exploiting highly sophisticated numerical codes like %CMBFAST  \cite{cmbfast}
%developed by U. Seljak and M. Zaldarriaga, or its supplementary versions  CMBEASY \cite{cmbeasy},
CAMB \cite{camb}. Of course, one has to rely on numericals at some point while performing data analysis from WMAP \cite{wmap3} but it is always better to see how far one can go analytically, which may sometime lead to deeper physical insight as well. This is the primary motivation for employing
semi-analytical treatment for perturbations in our typical model of inflation, viz. mutated hilltop inflation.
To this end, we mostly employ the framework of post-inflationary perturbations developed by Mukhanov (for an useful review see \cite{mukhanovb}) and obtain analytical results for CMB power spectrum for our typical model.
% 5 We shall further apply it to analyze  crucial physical phenomena such as Sachs-Wolfe effect and acoustic oscillation of the baryon-photon fluid at last scattering surface.
This will in turn help us reduce numerical complications to a great extent for our model.

%The plan of the paper is organized as follows: In the next section (section II) we have studied the theory of quantum fluctuations for mutated hilltop model
%of inflation and we have reproduced some results from our previous work \cite{barun} to make the present article comprehensive in nature. Section III deals with the post inflationary evolution of the perturbations for our model. In Section IV we have studied large scale CMB anisotropies mainly adiabatic Sachs-Wolfe effect and Section V deals with the acoustic oscillations of baryon-photon fluid, acoustic peaks and their dependence on various cosmological parameters in tight coupling approximation. Finally in section VI we end up with a discussion on the major outcomes of our work.

%%%%%%%%%%%%%%%%%%%%%%%%%%%%%%%%%%%%%%%%%%%%%%%%%%%%%%%%%%%%%%%%%%%%%%%%%%%%%%%%%%%%%%%%%%%%%%%%%%%%%%%%%%%%%%%%%%%%%%%%%%%%%%%%%%%%%%%%%%%%%%%%%%%%%
%%%%%%%%%%%%%%%%%%%%%%%%%%%%%%%%%%%%%%%%%%%%%%%%%%%%%%%%%%%%%%%%%%%%%%%%%%%%%%%%%%%%%%%%%%%%%%%%%%%%%%%%%%%%%%%%%%%%%%%%%%%%%%%%%%%%%%%%%%%%%%%%%%%%%%%

\section{Quantum fluctuation and origin of inhomogeneities}

The standard theory of cosmological perturbations is based on the formalism developed
in \cite{barun38}. %55 barun39,barun40}.
 To study the quantum fluctuations and subsequent classical perturbations
for mutated hilltop inflation let us directly refer to \cite{barun} where the inflationary potential has the following form
\begin{equation}\label{potential}
V(\phi)=V_0 \left[1- {\rm sech}(\alpha\phi) \right]
\end{equation}
This represents a single scalar field driven inflationary potential and falls in the wider class of hilltop
inflation \cite{hilllyth1}.

%%%%%%%%%%%%%%%%%%%%%%%%%%%%%%%%%%%%%%%%%%%%%%%%%%%%%%%%%%%%%%%%%%%%%%%%%%%%%%%%%%%%%%%%%%%%%%%%%%%%%%%%%%%%%%%%%%%%%%%%%%%%%%%%%%%%%%%%%%%%%%%%%%%%%%%%%%%%%%%%%%

\subsection{Curvature perturbation}
The expression for {\it comoving curvature perturbation} is acquired by solving the well known {\it Mukhanov-Sasaki} equation \cite{mukhanov1sasaki} and
in the slow-roll approximation we obtain
\beq\label{comcur}
{\cal R}_k\approx-\frac{\alpha\sqrt{V_0}}{M_P\sqrt{3}}\frac{|\eta|e^{-ik\eta}}{\sqrt{2k}}\left(1-\frac{i}{k\eta}\right)A(\eta)
\eeq
where $A(\eta)\equiv\ln
\left(a_1M_P^{-1}\sqrt{\frac{V_0}{3}} \vert\eta\vert \right)
$. Therefore the dimensionless spectrum for ${\cal R}_k$ turns out to be
\begin{equation}\label{ps}
P_{\cal R}(k)=\frac{\alpha^2V_0}{12\pi^2 M_P^2}(1+k^2\eta^2)A(\eta)^{2}
\end{equation}
%55 As discussed earlier, in our previous paper \cite{barun} we obtained the expression for
Evaluating the spectrum at horizon crossing, using the de-Sitter approximation $k=aH=-\eta^{-1}$ we obtain
\begin{equation}\label{psds}
P_{\cal R}(k)|_{k=aH}=\frac{\alpha^2V_0}{6\pi^2 M_P^2}A(\eta)^{2}
\end{equation}
%where $\eta_{hc}$ is conformal time at horizon crossing.
But in adopting exact de-Sitter relation at horizon crossing we are discarding the effect of the evolution of
the scalar field. In what follows we shall pursue a non trivial path.  We first note
that for the inflation model under consideration at the time of horizon crossing,
\begin{equation}\label{ah}
1+k^2\eta^2\approx2-\frac{1}{\alpha^2\sqrt{\frac{V_0}{3}}M_P(d-t)}
\end{equation}
which is slightly different from the usual de-Sitter relation
$1+k^2\eta^2=2$. Eqn.\eqref{ah} shows that there will be direct effect of the evolution
of the scalar field on the observable parameters when they are evaluated at horizon exit. Our approach is a bit
similar to \cite{peiris}, where variable speed of the scalar field fluctuations have been taken into account for evaluating
observable quantities at horizon exit. Following this trail the power spectrum turns out to be
\begin{equation}\label{pshc}
P_{\cal R}|_{k=aH}=\frac{\alpha^2V_0}{12\pi^2
M_P^2}\left[2-\frac{\alpha^{-2}M_P^{-2}}{A(\eta)}\right]A(\eta)^2
\end{equation}
Comparison of the Eqns.(\ref{psds}) and (\ref{pshc}) shows that the correction term indeed has some
significant role from theoretical aspects. We may now re-derive all the allied expressions for the observable quantities in a more
accurate way. The expression for the scalar spectral index %\cite{lythc}
is now given by
\beq n_s=1-\left[\frac {4A(\eta)-\alpha^{-2}M_P^{-2}}{2A(\eta)^2-\alpha^{-2}M_P^{-2}A(\eta)}\right]
\eeq
%5 Moreover, we know from the CMB fluctuations the power spectrum of the comoving curvature
%perturbation has the observational bound  \cite{cmb} as $P_{\cal
%R}^{1/2} \sim 5 \times 10^{-5}$.In due course, the estimations of these observable parameters will be performed in the present framework.
We have also succeeded in calculating the running of the spectral index which turns out to be
\begin{equation}
\displaystyle \frac{dn_s}{d \ln k}{\displaystyle \left|_{k=aH}
\right.}=-\left[ \frac{8A(\eta)^2-4\alpha^{-2}M_P^{-2}A(\eta)+\alpha^{-4}M_P^{-4}}{\left(2A(\eta)^2-\alpha^{-2}M_P^{-2}A(\eta)\right)^2}\right]
\end{equation}
Clearly, though the quantity within the parentheses is quite small still gives a nonzero value resulting in a negative running of the spectral index
consistent with WMAP3 data set.

%%%%%%%%%%%%%%%%%%%%%%%%%%%%%%%%%%%%%%%%%%%%%%%%%%%%%%%%%%%%%%%%%%%%%%%%%%%%%%%%%%%%%%%%%%%%%%%%%%%%%%%%%%%%%%%%%%%%%%%%%%%%%%%%%%%%%%%%%%%%%%%%%%%%%%%%%

\subsection{Tensor fluctuation}

%The formulation of quantum theory of fluctuation in inflationary
%cosmology is incomplete unless we discuss on the possibility of
%generation of  primordial gravitational waves from tensor
%fluctuations.

The tensor modes representing primordial gravitational waves,
obtained in a similar way as the scalar modes and the corresponding spectrum is given by
\begin{equation}\label{ptf}
P_{T}|_{k=aH}=\frac{V_0}{3\pi^2M_P^4}\left[2-\frac{\alpha^{-2}M_P^{-2}}{A(\eta)}\right]
\end{equation}
%Indeed, this is the  expression for the primordial power spectrum
% of the gravitational waves. % Detection of these waves would, quite
%remarkably, measure the energy scale associated with inflation.
As expected, the spectrum for the tensor modes is also modified in this approach. The corresponding spectral
index turns out to be
\begin{eqnarray}\label{nt}
n_T&=&-\frac{1}{\alpha^2M_P^2}\left[\frac{1}{2A(\eta)^2-\alpha^{-2}M_P^{-2}A(\eta)}\right]
\end{eqnarray}

From Eqns.(\ref{ptf}) and (\ref{pshc}) we arrive at the expression for
the ratio of tensor to scalar amplitudes, given by
\begin{eqnarray}\label{r}
 r&=& 4\frac{\alpha^{-2}M_P^{-2}}{A(\eta)^2}
\end{eqnarray}
We shall use this expression in calculating $r$
and subject it to  observational verification ($r < 0.1$) \cite{wmap07}.

Combining Eqns.(\ref{r}) and (\ref{nt}) we obtain a vital relation between tensor spectral index $n_T$ and the tensor to
scalar amplitude ratio $r$ as,
\begin{equation}\label{consistency}
r = -8 n_T\left(1 - \frac{\sqrt r}{4\alpha M_P}\right)
\end{equation}
The spectrum $P_{T_{|_{k=aH}}}$ of tensor  perturbation
conveniently specified by the tensor fraction
$r=\frac{P_{T}{|_{k=aH}}}{P_{\cal R}{|_{k=aH}}}$ yields the
relation $r=-8n_T$ in the slow-roll approximation
\cite{andrew,david1}. But Eqn.(\ref{consistency}) shows that when the explicit effect of the scalar field evolution is taken into account in evaluating the observable parameters at horizon exit we obtain a consistency relation which is slightly modified. Of course, there exist in the literature other ways of obtaining a modified consistency relation. Such a modified consistency relation can be found in any analysis where higher order terms in the expansion
of slow roll parameters are taken into account \cite{0101225,PRD66}. The consistency relation is also modified in the context of brane inflation \cite{peiris,bean} and non-standard models of inflation \citep{mukhanov} where generalized propagation speed (less than one) of the scalar field fluctuations relative to the homogeneous background have been considered. Further, deviation from the usual consistency relation can be found in \citep{parker} where {\it tensor to scalar ratio} has been shown to be a function of {\it tensor spectral index, scalar spectral index} and {\it running of the tensor spectral index}. Our approach is somewhat similar to these.

Finally,  we estimate the observable parameters from
the first principle of the theory of fluctuation as derived in this
section for three sets of values of $\alpha$ and tabulate it in
Table \ref{tab2}.
\begin{table}
\begin{tabular}{|c|c|c|c|c|}
\hline $\alpha $ & $a_1$  & $P_R^{1/2}$&$n_{s}$& $r$ \\
$M_{P}^{-1}$&$M_{P}^{-1}$ &  & &  \\
 \hline $2.9$& $6.7091\times10^{25} $
 &$3.77624\times10^{-5}$&0.96088 &$ 1.81759\times10^{-4}$\\
\hline $3.0$& $6.6492\times10^{25} $
 &$3.90592\times10^{-5}$&0.96088 &$ 1.69903\times10^{-4}$\\
\hline $3.1$& $6.5945\times10^{25} $
 &$4.03766\times10^{-5}$&0.96087 &$ 1.59170\times10^{-4}$\\
 \hline
\end{tabular}
\caption{Table for the observable quantities as obtained from the theory of fluctuations}
\label{tab2}
\end{table}
 From the table it is quite clear that the observable parameters related to
 perturbations, %5555 {\em viz}, $P_{\cal R}^{1/2}$, $n_{s}$ and $r$,  calculated from the theory of
% quantum fluctuations,
are in excellent agreement with observational data.

%%%%%%%%%%%%%%%%%%%%%%%%%%%%%%%%%%%%%%%%%%%%%%%%%%%%%%%%%%%%%%%%%%%%%%%%%%%%%%%%%%%%%%%%%%%%%%%%%%%%%%%%%%%%%%%%%%%%%%%%%%%%%%%%%%%%%%%%%%%%%%%%%%%%%%%%%
%%%%%%%%%%%%%%%%%%%%%%%%%%%%%%%%%%%%%%%%%%%%%%%%%%%%%%%%%%%%%%%%%%%%%%%%%%%%%%%%%%%%%%%%%%%%%%%%%%%%%%%%%%%%%%%%%%%%%%%%%%%%%%%%%%%%%%%%%%%%%%%%%%%%%%%%%
\section{Post-inflationary evolution of perturbations}
Now to relate the initial perturbation seeds with cosmological observables we study post-inflationary evolution of these perturbation proceeding analytically as far as possible from our specific model.

% 5 The fluctuation $\Phi$ of the gravitational potential governs both the matter density fluctuation $\delta_m$ and temperature fluctuation $\frac{\delta T}{T}$
% in the CMB.
While a perturbation mode re-enters the horizon during matter dominated era, we obtain the approximate solution for the fluctuations in the gravitational potential from the relation
\begin{eqnarray}\label{sphi}
\Phi_k&\approx&-\frac{3}{5} {\cal R}_k
\end{eqnarray}
It should be pointed out that Eqn.\eqref{sphi} is an approximate solution for the gravitational potential so that we can
neglect contributions from smaller scales spatial gradient. Still qualitative information about the evolution of $\Phi$ in the large scale regime is
transparent. Using the near constancy of ${\cal R}_k$ in the superhorizon scales \cite{malik},
we obtain an approximated expression for the comoving curvature perturbation and for our typical model %, (using $k=aH=\frac{2}{\eta}$),
at horizon reentry during matter domination this turns out to be
\begin{equation}\label{supercomcur}
{\cal R}_k\simeq\frac{\alpha}{M_P}\sqrt{\frac{V_0}{3}}
\frac{i}{\sqrt{2k^3}}B(k)
\end{equation}
where $B(k)\equiv\ln\left(2 a_1M_P^{-1}{\sqrt\frac{V_0}{3}}k^{-1}\right)$.
% 5 The scale dependence of ${\cal R}_k$ makes $\Phi$, the fluctuation in the gravitational potential a scale dependent quantity, albeit slightly.
%Combining the Eqns.(\ref{sphi}) and (\ref{supercomcur}), the expression for the
%$k^{th}$ Fourier mode of
So the fluctuation modes of the gravitational potential, entering the horizon during the matter dominated era, are given by,
\begin{equation}\label{phi}
\Phi|_{k=aH}\approx-\frac{3\alpha}{5M_P}\sqrt{\frac{V_0}{3}}
\frac{i}{\sqrt{2k^3}}B(k)
\end{equation}
From the above expression it is apparent that in this model the gravitational
potential is not strictly scale invariant, but has a slight scale
dependence with a mild running as expected.

%5 When the curvature perturbation re-enters the horizon during matter
%dominated era they create density fluctuations $\delta\rho_k$
%through the gravitational attractions of the potential wells.
Using Eqn.\eqref{phi}, the spectrum of the matter density contrast can be found from Poisson equation, turns out to be
\begin{equation}\label{gtrpowhor}
P_\delta|_{k=aH}\approx\frac{16 \alpha^2 V_0}{75\pi^2M^2_P}B(k)^2
\end{equation}
and the corresponding spectral index and the running are respectively given by
\beq\label{gtrspectralindex}
n_\delta\approx 1-2B(k)^{-1},~n^{'}_\delta\approx-2B(k)^{-2}%\label{gtrspectralindex}\\
%\label{gtrspectralrun}
\eeq
Eqn.(\ref{gtrspectralindex}) shows that the spectral index is not exactly scale invariant, it has a small negative tilt with negative running.% given by Eqn.\eqref{gtrspectralrun}.

The fluctuations produced during inflation are scale dependent as is evidenced in Section II and this is reflected in the matter spectrum.
 As a consequence, we obtain negative scalar spectral tilt and negative running of the scalar spectral index for the matter power spectrum for our model.
  %%555 This reflects direct impact of employing semi-analytical treatment for mutated hilltop model of inflation.

%%%%%%%%%%%%%%%%%%%%%%%%%%%%%%%%%%%%%%%%%%%%%%%%%%%%%%%%%%%%%%%%%%%%%%%%%%%%%%%%%%%%%%%%%%%%%%%%%%%%%%%%%%%%%%%%%%%%%%%%%%%%%%%%%%%%%%%%%%%%%%%%%%%%%%%%
%%%%%%%%%%%%%%%%%%%%%%%%%%%%%%%%%%%%%%%%%%%%%%%%%%%%%%%%%%%%%%%%%%%%%%%%%%%%%%%%%%%%%%%%%%%%%%%%%%%%%%%%%%%%%%%%%%%%%%%%%%%%%%%%%%%%%%%%%%%%%%%%%%%%%%%%
\section{Temperature anisotropies in CMB }%and Sachs-Wolfe effect}

Let us now proceed further with the analytical treatment and discuss the CMB angular power spectrum resulting from temperature anisotropies due to scalar curvature perturbations for our typical inflationary model. The dominant contribution to the large scale CMB anisotropy comes from the Sachs-Wolfe effect  \cite{sachs}. In the sudden decoupling approximation for adiabatic perturbation, the assumption of complete matter domination at last scattering provide us the
Sachs-Wolfe spectrum for the typical model of our consideration, which is approximately given by
\begin{eqnarray}\label{spectrum1}
C^{\mbox{SW}}_l\approx\frac{\alpha^2V_0}{75\pi M^2_P}
\left(\ln\left[a_1M^{-1}_P\sqrt{\frac{V_0}{3}}\frac{2\eta_0}{l}\right]
\right)^2\frac{1}{2l(l+1)}
\end{eqnarray}
Here $\eta_0$ is the present dimensionless comoving distance to the last scattering surface.

\begin{figure}
\centerline{\includegraphics[width=7cm, height=4cm]{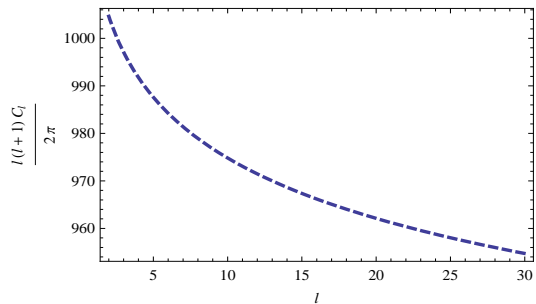}}
  \caption{\label{sachswolfeplateau} Sachs-Wolfe plateau}
\end{figure}
Figure \ref{sachswolfeplateau} shows variation of angular power spectrum for Sachs-Wolfe effect $\frac{l(l+1)}{2\pi}C^{SW}_l$ with the CMB multipoles $l$ in units of $\mu K^2$, where we have taken the following representative values for the quantities involved: $V_0^{1/4}=1.4\times10^{-3}M_P$, $ \alpha=3.0M_P^{-1}$,
$a_1=6.6492\times10^{25}M_P^{-1}$ and $\eta_0=7.42438\times10^{26}$. Figure \ref{sachswolfeplateau} reveals that the Sachs-Wolfe plateau is not exactly flat but is slightly tilted towards larger values of $l$. This is not surprising, since the primordial curvature perturbation in mutated hilltop inflation
is not strictly scale-invariant. %5 However, as of now this is an interesting result which reveals the credentials of analytical calculations of post-inflationary
%5 perturbations for this type of inflationary models.
%%%%%%%%%%%%%%%%%%%%%%%%%%%%%%%%%%%%%%%%%%%%%%%%%%%%%%%%%%%%%%%%%%%%%%%%%%%%%%%%%%%%%%%%%%%%%%%%%%%%%%%%%%%%%%%%%%%%%%%%%%%%%%%%%%%%%%%%%%%%%%%%%%%%%%%%%
%%%%%%%%%%%%%%%%%%%%%%%%%%%%%%%%%%%%%%%%%%%%%%%%%%%%%%%%%%%%%%%%%%%%%%%%%%%%%%%%%%%%%%%%%%%%%%%%%%%%%%%%%%%%%%%%%%%%%%%%%%%%%%%%%%%%%%%%%%%%%%%%%%%%%%%%%%
%5 \section{Baryon acoustic oscillation and CMB multipoles}

In the smaller scales CMB spectrum is dominated by acoustic oscillation of the baryon-photon fluid. The
approximate solution to the acoustic oscillation equation \cite{lythR, liddlelyth}%, hu, white}
for our case turns out to be
\begin{equation}\label{finalacos}
\frac{1}{4}\delta_{\gamma k}\approx-(1+R)\Phi_k+\frac{1}{2}T^0_k\Phi^0_k
\cos(kr_s)
\end{equation}
where $r_s\equiv\int^{\eta}_0 c_s(\eta)d\eta$, $c_s$ is the sound speed, $R=\frac{3\rho_b}{4\rho_{\gamma}}$, $\Phi_k\equiv-\frac{3}{5}T_k {\cal R}_k$ is the small scale solution of the gravitational potential with $T_k$ being corresponding transfer function and we have defined
\begin{equation}\label{priphi}
\Phi^0_k\equiv-\frac{2\alpha}{3M_P}\sqrt{\frac{V_0}{3}}\frac{1}{\sqrt{2k^3}}\left[\ln
\left(a_1M_P^{-1}{\sqrt\frac{V_0}{3}}k^{-1}\right)\right]
\end{equation}
to be the initial gravitational potential fluctuation and $T^0_k$ is the associated transfer function.
The expression for the photon velocity perturbation $V_{\gamma k}$ is then given by
%5555, we differentiate (\ref{finalacos}) w.r.t. conformal time and use the relation,
%$\frac{1}{4}{\delta^{'}_{\gamma k}}=-\frac{k}{3}V_{\gamma k}+\Phi^{'}_k$, we find
\begin{equation}\label{phovel}
V_{\gamma k}\approx\frac{3c_s}{2}T^0_k\Phi^0_k\sin(kr_s).
\end{equation}
%%55555  Now including the Silk damping effect \cite{silk} and finite thickness of the last scattering surface,
So the CMB angular power spectrum for our typical model turns out to be \cite{mukhanovb}
\begin{equation}\label{cmbspectrum1}
C_{l}\approx4\pi P_{\Phi^0}\int^{\infty}_0 \left[C-D\cos(\rho l
x)+E\cos^2(\rho l x)+F\left(1-\frac{l(l+1)}{l^2
x^2}\right)\sin^2(\rho l x)\right]j^2_l(x l)\frac{dx}{x}
 \end{equation}
here $x=\frac{k\eta_0}{l}$, and the functions in the integrand are defined as follows:
\bea C&\equiv&
\frac{81}{100}R^2{T}^2(x)e^{-\frac{l^2x^2}{l^2_f}}, ~D\equiv \frac{9}{10}RT(x)T^0(x)e^{-1/2\frac{l^2x^2}{l^2_f+l^2_s}},~
E\equiv \frac{1}{4}{T^{0}}^2(x)e^{-\frac{l^2x^2}{l^2_s}},~F\equiv\frac{9}{4} c^2_s {T^{0}}^2(x)e^{-\frac{l^2x^2}{l^2_s}},\nonumber\\~l^{-2}_f&\equiv& 2\sigma^2(\frac{\eta_{LS}}{\eta_0})^2,~l^{-2}_s\equiv 2\left[\sigma^2+(k_D\eta_{Ls})^{-2}\right](\frac{\eta_{LS}}{\eta_0})^2,~\rho\equiv \frac{1}{\eta_0}\int^{\eta_{LS}}_0 c_s d\eta,~\sigma\equiv 1.49\times 10^{-2}\left[1+\left(1+z_{EQ}/z_{LS}\right)^{-1/2}\right]\nonumber
\eea
with $ z_{EQ}~\mbox{ and }~z_{LS}$ are redshifts corresponding to matter-radiation equality and last scattering surface respectively and
$k^{-2}_D\equiv\left[\frac{2}{5}\int^{\eta}_0 c^2_s \frac{\tau_\gamma}{a} d\eta\right]$, is the Silk damping scale \cite{silk,liddlelyth,dodelson}.
Here $\tau_\gamma$ is the mean free time for photon scattering and ``LS" stands for last scattering.
%%5555 We have also used the well known result $\Phi^{0}_k=\frac{10}{9}\Phi|_{\eta=\eta_{LS}}$. The primordial power spectrum of the gravitational potential, defined as
%$P_{\Phi^0}$ below, is a characteristic feature of the inflation
%model under consideration that is reflected on CMB angular power
%spectrum as a direct imprint  of a typical model of inflation.
%Consequently, in our model,
The \textit{slight} scale dependence of the primordial curvature perturbation is reflected here via
\begin{equation}
P_{\Phi^0}\equiv\frac{\alpha^2V_0}{27\pi^2 M^2_P}\left[\ln\left(a_1M_P^{-1}{\sqrt\frac{V_0}{3}}~\frac{\eta_0}{l}\right)\right]^2
\end{equation}
%5555 Taking large $l$ limit of the spherical Bessel function,
%Eqn.(\ref{cmbspectrum1}) can be written as
%\begin{equation}\label{cmbspectrumlargel}
%C_{l}\approx\frac{2\pi}{l^2}P_{\Phi^0}\int^{\infty}_1
%\left[\frac{C-D\cos(\rho l x)+E\cos^2(\rho l
%x)+F\left(1-\frac{l(l+1)}{l^2 x^2}\right)\sin^2(\rho l
%x)}{x^2\sqrt{x^2-1}}\right]dx
%\end{equation}
The results we have obtained may not be exact, but still qualitative
behavior of the associated physical quantities can be extracted
from them very easily.

%\begin{figure}[htb]
%\centerline{\includegraphics[width=7cm, height=5cm]{peak.eps}}
% \caption{\label{peak} Acoustic Oscillations of Baryon-Photon Fluid at Last Scattering Surface
%}
%\end{figure}
\begin{figure}[htb]
\centerline{\includegraphics[width=7cm, height=5cm]{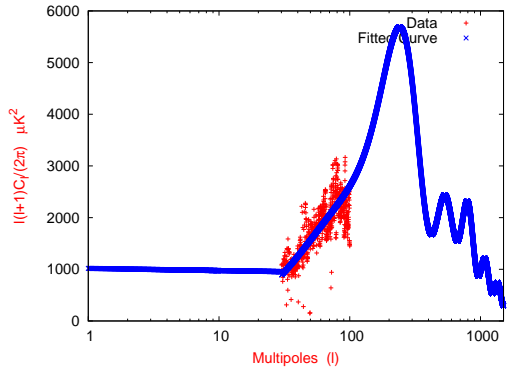}}
 \caption{\label{cmb} The total CMB angular power spectrum for adiabatic modes}
\end{figure}
%In Figure \ref{peak} we have plotted the variation of temperature
%anisotropy spectrum with the multipoles $(l)$ taking the large
%$l(\geq100)$ limit which depicts true characteristics of baryon acoustic oscillation in CMB.

To analyze the entire CMB angular power spectrum, we shall %rather
concentrate on Figure \ref{cmb} which depicts its behavior for the entire significant range of the multipoles ($1\leq l\leq1500$) and
combines \textit{almost} all the effects arising in CMB. Figure \ref{cmb} can be split into 3 regions: (i) $1\leq l\leq30$
represents mainly Sachs-Wolfe effect, (ii) for $30<l\leq100$ we have considered the exact expression for the Bessel function in
Eqn.(\ref{cmbspectrum1}) and finally (iii) for $100<l\leq1500$ large $l$ limit of the Bessel function has been used %using Eqn.(\ref{cmbspectrumlargel})
for a rigorous calculation of the CMB spectrum. For the plots we have used the  transfer functions
from \cite{mukhanovb} with appropriate modifications: $T^0(x)=1.20+0.09\ln\left(\frac{I_\Lambda l~x}{250\sqrt{\Omega_m}}\right)$ and
$T(x)=0.52-0.21\ln\left(\frac{I_\Lambda l~x}{250\sqrt{\Omega_m}}\right)$; where the late time effects due
to dark energy have been incorporated via $I_\Lambda$ which is defined as $I_\Lambda\equiv 3\left(\frac{\Omega_\Lambda}{\Omega_M}\right)^\frac{1}{6}\left[\int^{\sinh^{-1}
\sqrt{\Omega_\Lambda/\Omega_M}}_0  \frac{dx}{(\sinh
x)^{2/3}}\right]^{-1}$.
%55555 The final plot has been drawn by
%generating data points by numerical integration of the functions
%in Eqn.(\ref{cmbspectrum1}) and Eqn.(\ref{cmbspectrumlargel}) with
%appropriate limits and using gnuplot.
%55555 The integral is evaluated
%with the following set of numerical values of the parameters
%involved: $h=0.71, \Omega_B=0.0449, \Omega_M=0.2669,
%\Omega_\Lambda=0.734, \Omega_\gamma h^2=2.48 \times 10^{-5},
%z_{LS}=1090, z_{EQ}=3300, R_{LS}=0.627405,
%\sigma=2.23245\times10^{-2},  l_f=1617, l_s=1254,
%I_\Lambda=1.11726,  \rho=0.0112647, \eta_0=7.42438\times10^{26}$.
%Further, here we have assumed that radiation domination starts
%immediately after the end of inflation  at a redshift around
%$z\approx3.3\times10^8$ to get the present value of the
%dimensionless conformal time.

%%%55555 As is well known, the positions and heights of the peaks provide crucial information
%about different cosmological parameters.
% The peak positions depend strongly on the  parameter
%$\rho$ and hence on the  baryon to photon ratio at last
%scattering. Thus, the heights of the peaks are very sensitive to
%the baryon fraction. Also the peak positions (more specifically,
%the position of the first peak) are sensitive to the curvature of
%the space and on the rate of cosmological expansion hence on the
%dark energy and other forms of the matter. The second and third
%peaks constraint dark matter and baryonic matter densities of the
%universe.
In Figure \ref{cmb} resulting from our analysis the first and the most prominent peak of the acoustic oscillation
arises at $l\approx241$ and at a height of $\approx 5900 \mu K^2$.
%55 This is in accordance with observational outcome that our universe
%is spatially flat (with spatial curvature $k = 0$) and also, the dark energy density fraction of the universe is nearly $70 \%$ of the total cosmic density.
The first peak is followed by two nearly equal height peaks at $l\approx533$ and $ 791$.
%These conform with the estimated values of dark matter and baryonic matter densities. Also, the positions of the peaks appear at
%nearly integral multiples of the first peak, which is a direct outcome of our analysis in which we consider adiabatic
%perturbations and use corresponding initial conditions a priori.
After the third peak there is a damping tail in the oscillation with successive peaks with lower heights. This happens due to considering
Silk damping effect in our analysis, which is also consistent with CMB observations.
%As mentioned, for $30<l\leq100$ we have considered the exact expression for Bessel
%function in Eqn.(\ref{cmbspectrum1}). This feature has been reflected in the increased error bars within this region (compared
%to rest of the range of $l$) in generating data points.
So mutated hilltop inflation conform well with the recent observations.

%%%%%%%%%%%%%%%%%%%%%%%%%%%%%%%%%%%%%%%%%%%%%%%%%%%%%%%%%%%%%%%%%%%%%%%%%%%%%%%%%%%%%%%%%%%
 \section{Summary and Discussions}

 In this article we have studied quantum fluctuations and corresponding
classical perturbations  as well as allied observational aspects by a semi-analytical treatment,
based on mutated hilltop model of inflation. We have succeeded in deriving analytical expressions for
most of the observable parameters. An approximated expression for the fluctuations in the gravitational fluctuations has
also been successfully obtained using relativistic perturbations. We have also obtained an expression for the scale dependent matter power
spectrum, along with a negative running of the corresponding spectral index which are in accord with recent observational demands. We have shown explicitly that in mutated hilltop inflation the Sachs-Wolfe plateau is not strictly flat but is slightly tilted towards the larger
multipoles. Moreover, we have also studied the baryon acoustic oscillation based on our model by carrying the analytical framework as far as
possible, which gives physical insight to the scenario and reduces numerical complications to a great extent. Finally, we have
employed certain simplistic numerical techniques and found that the positions of the acoustic peaks conform  well within the
estimated values of different cosmological parameters.

Some comments on employing this semi-analytical treatment to mutated hilltop inflation model are in order. The resulting
%%This approach indeed gives rise to significant qualitative features in
CMB power spectrum as obtained from our analysis is quite consistent with what can be obtained from direct numerical techniques
via CAMB \cite{camb}. Of course, using more complicated numerical technique via Monte-Carlo simulation like COSMOMC \cite{cosmomc} will result in
more accurate estimation of individual cosmological parameters but, at a first go, this semi-analytical approach is more or less sufficient for
extracting physics out of our typical model. In future we would like to employ COSMOMC to our model as a complimentary techniques of extracting
physics out of mutated hilltop inflation model.

\section*{Acknowledgments}
SP thanks members of Centre for Theoretical Studies, IIT Kharagpur and Physikalisches Institut, Universit\"{a}t Bonn for illuminating discussions.
BKP thanks Council of Scientific and Industrial Research, Govt. of India for
financial support through Junior Research Fellowship (Grant No.
09/093 (0119)/2009). The work of SP is supported by a research grant from Alexander von Humboldt Foundation, Germany, and  is
partially supported by the SFB-Tansregio TR33 ``The Dark
Universe'' (Deutsche Forschungsgemeinschaft) and the European
Union 7th network program ``Unification in the LHC era''
(PITN-GA-2009-237920).

%%%%%%%%%%%%%%%%%%%%%%%%%%%%%%%%%%%%%%%%%%%%%%%%%%%%%%%%%%%%%%%%%%%%%%%%%%%%%%%%%%%%%%%%%%%%%%%%%%%%%%%%%%%%%%%%%%%%%%%%%%%%%%%%%%%%%%%%%%%%%%%%%%%%%%%%%%


\begin{references}
\bibitem{guth} A. H. Guth, Phys. Rev. D{\bf 23}, 347 (1981)
\bibitem{wmap3} M. Tegmark {\it et. al.}, Phys. Rev. D {\bf 69}, 103501 (2004)
\bibitem{wmap07} WMAP collaboration, D. N. Spergel et al.,  Astrophys. J.  Suppl. {\bf 170}, 377 (2007);
for uptodate results on WMAP, see
{\em http://lambda.gsfc.nasa.gov/product/map/current}
\bibitem{large} T. Padmanabhan, {\it Theoretical Astrophysics}, Volume III: {\it Galaxies and Cosmology}, Cambridge University Press, U. K. (2002)

%\bibitem{5} A. R. Liddle and D. H. Lyth, {\it Cosmolgical Inflation and Large Scale Structure}, Cambridge University Press, U. K. (2000)
%\bibitem{6} A. Albrecht and P. J. Steinhardt, Phys. Rev. Lett. {\bf 48}, 1220 (1982)
%\bibitem{8} A. D. Linde, Phys. Lett B {\bf 108}, 389 (1982)
%\bibitem{10} S. Kachru, R. Kallosh, A. Linde and S. P. Trivedi,  Phys. Rev. D
%{\bf 68}, 046005 (2003)
%\bibitem{11} M. Dine, L. Randall and S. Thomas, Phys. Rev. Lett. {\bf 75}, 398 (1995)
%\bibitem{15} S. Pal, arXiv: 0808.1630 [gr-qc]

%\bibitem{penzias} A. A. Penzias and R. W. Wilson, Astrophys. J. {\bf 142}, 419 (1965)
%\bibitem{julian67} W. H. Julian, Astrophys. J. {\bf 148} 175 (1967)
%\bibitem{jles07} J. Lesgourgues, M. Viel, M.G. Haehnelt and R. Massey, JCAP {\bf 0711}, 008 (2007)
\bibitem{barun} B. K. Pal, S. Pal and B. Basu, JCAP {\bf 1001}, 029 (2010)
\bibitem{andrew} A. R. Liddle and D. H. Lyth, Phys. Lett. B {\bf 291}, 391 (1992)
\bibitem{stewart} E. D. Stewart and D. H. Lyth Phys. Lett. B {\bf 302}, 171 (1993)
\bibitem{0101225} Ewan D. Stewart and Jin-Ook Gong, Phys. Lett. B {\bf510}, 1 (2001)
\bibitem{PRD66}Eckehard W. Mielke and Humberto H. Peralta,  Phys. Rev. D {\bf 66}, 123505 (2002)
\bibitem{parker} I. Agullo et al. Gen. Rel. Grav. {\bf 41}, 2301 (2009)
\bibitem{peiris} H. V. Peiris et al., Phys. Rev. D{\bf 76}, 103517 (2007)
\bibitem{bean}  R. Bean et al., JCAP {\bf 0705} 004 (2007)
\bibitem{mukhanov} J. Garriga and V. F. Mukhanov, Phys. Lett. B{\bf 458}, 219 (1999)

\bibitem{cobe} G. F. Smoot et al., Astrophys. J. {\bf 396}, L1 (1992); S. Dodelson  and J. M. Jubas, Phys. Rev. Lett. {\bf 70}, 2224 (1993)
\bibitem{planck} Planck collaboration, {\em http://www.rssd.esa.int/index.php?project=Planck}; P. A. R. Ade et.al., arxiv:1101.2022

%\bibitem{cmbfast} Online link: \textit{http://lambda.gsfc.nasa.gov/toolbox/tb\_cmbfast\_ov.cfm}
%\bibitem{cmbeasy} Online link: \textit{http://www.thphys.uni-heidelberg.de/~robbers/cmbeasy/}
\bibitem{camb} Online link: \textit{http://camb.info/}
\bibitem{mukhanovb} V. F. Mukhanov, {\it Physical Foundation of Cosmology}, Cambridge University Press, U.K.(2006)
\bibitem{barun38}V. F. Mukhanov and G. V. Chibisov, JETP Lett. {\bf 33}, 532 (1981);
S. W. Hawking, Phys. Lett. B {\bf 115}, 295 (1982); A. A.
Starobinsky, Phys. Lett. B {\bf 117}, 175 (1982); A. H. Guth and S.
Y. Pi, Phys. Rev. Lett. {\bf 49}, 1110 (1982)
%\bibitem{barun39} D. H. Lyth and D. Wands, Phys. Lett. B {\bf 524}, 5(2002); D. H. Lyth, C. Ungarelli and D. Wands, Phys. Rev. D {\bf67}, 023503(2003)
%\bibitem{barun40}D. Langlois, arXiv: hep-th/0405053; D. Langlois, arXiv: 0811.4329 [gr-qc]
\bibitem{hilllyth1}L. Boubekeur and D. H. Lyth, JCAP {\bf 0507},   010 (2005); K. Kohri, C.-M. Lin and D. H. Lyth, JCAP {\bf 0712}, 004 (2007)
%\bibitem{hilllyth2}
\bibitem{mukhanov1sasaki} V. F. Mukhanov, JETP Lett. {\bf 41}  493 (1986); M. Sasaki, Prog. Theor. Phys. {\bf 76}, 1036 (1986)
%\bibitem{sasaki}

%\bibitem{bd} T. S. Bunch and P. C. W. Davies, Proc. Roy. Soc. Ser. A {\bf 360} 117 (1978)

%\bibitem{lythc} D. H. Lyth, JCAP {\bf 0511}, 006 (2005)
%\bibitem{cmb} D. N. Spergel et al., Astrophys. J. {\bf 148}, 175 (2003)

\bibitem{david1} A. R. Liddle and D. H. Lyth, Phys. Rep. {\bf 231}, 1 (1993)
%\bibitem{hj} B. K. Pal et al., arXiv:1105.6362

\bibitem{malik} D. H. Lyth, K. A. Malik, M. Sasaki, JCAP {\bf 0505} 004 (2005)
\bibitem{sachs} R. K. Sachs and A. M. Wolfe, Astrophys. J. {\bf 147}, 73 (1967)
\bibitem{lythR} D. H. Lyth, Phys. Rev. D{\bf 31}, 1792(1985); W. Hu and M. Sugiyama, Astrophys. J. {\bf 471}, 542 (1996);
\bibitem{liddlelyth}D. H. Lyth and A. R. Liddle,  {\it The Primordial Density Perturbation}, Cambridge University Press, U.K.(2009)
%\bibitem{hu} W. Hu and M. Sugiyama, Astrophys. J. {\bf 471}, 542 (1996)
%\bibitem{white} W. HU and M. White, Astrophys. J. {\bf 471}, 30 (1996)
\bibitem{silk} J. Silk, Nature {\bf 215}, 1155 (1972)
\bibitem{dodelson} S. Dodelson, {\it Modern Cosmology, Academic Press, USA (2003)}
%\bibitem{step} D. K. Hazra, M. Aich, R. K. Jain, L. Sriramkumar and T. Souradeep, JCAP \textbf{1010}, 008,2010
\bibitem{cosmomc} http://cosmologist.info/cosmomc/
\end{references}
\end{document}